\documentclass{article}
\usepackage{spconf,amsmath,graphicx}

\title{Understanding Audio Pattern Using Convolutional Neural Network from Raw Waveforms}
\name{Shuhui Qu*, Juncheng Li*, Wei Dai, Samarjit Das}
\address{shuhuiq@stanford.edu, billy.li@us.bosch.com, wdai@cs.cmu.edu, samarjit.das@us.bosch.com}
\begin{document}
%
\maketitle
\begin{abstract}
One key step in audio signal processing is to transform the raw signal into representations that are efficient for encoding the original information. Traditionally, people transform the audio into spectral representations, as a function of frequency, amplitude and phase transformation. In this work, we take a purely data-driven approach to understand the temporal dynamics of audio at the raw signal level. We maximize the information extracted from the raw signal through a deep convolutional neural network (CNN) model. Our CNN model is trained on the \emph{urbansound8k} dataset~\cite{salamon2014dataset}. We discover that salient audio patterns embedded in the raw waveforms can be efficiently extracted through a combination of nonlinear filters learned by the CNN model.
\end{abstract}
\section{Introduction}
\label{sec:intro}

In audio signal processing and analysis, one key step is to transform the raw signal into representations that are suitable for the task at hand e.g., audio pattern classification.~\cite{smith2006efficient}\cite{smith2004learning}. Traditional studies describe the auditory representations in terms of spectral features, such as frequency, amplitude and phase transformation. Rather than merely analyzing audio's impulse response, there is existing effort trying to study the entire generation and origination of audio signals as a system. In ~\cite{smith2006efficient}\cite{smith2004learning}, Smith and Lewicki applied dictionary learning to identify efficient auditory codes. They found that these auditory codes show similarities to time-domain cochlear filter estimates, and have a frequency-bandwidth dependence similar to that of auditory nerve fibres. Recent advancement of high-performance computing hardwares and more available large datasets enabled us to utilize deeper models such as CNNs on raw waveform and maximize the information extracted. 

Several attempts have been made to employ CNNs for feature extraction from the raw signal in end-to-end speech recognition domain~\cite{golik2015convolutional}\cite{sainath2013deep}\cite{tuske2014acoustic}\cite{palaz2015analysis}, and these methods obtained good accuracy. These studies have shown that the learned weight of first layer has the pattern as a set of narrow bandpass filters.

In this study, we also apply the CNNs to the raw waveform of environmental sound \emph{urbansound8k}~\cite{salamon2014dataset} and understand the mechanism of the convolutional layer, especially the first layer. We first compare the mechanism between wavelet transformation and convolutional layers. After running experiments, we verify that the learned kernels operate similarly to band pass filters. We also take the inverse of these kernels to reconstruct original signal, and the reconstructed signal captures the pattern of the original signal and has minimum loss compared with the original signal. This indicates that the data driven method could adaptively, efficiently approximate the fundamental properties of audio as a time-domain signal.

The rest of the paper is structured as the followings: The relation between wavelet transformation and convolutional layers are introduced in section 2. Experimental setup is described in section3. Result and analysis are shown in section 4. Conclusions are drawn in section 5.

\section{Methodology}
Traditionally, Fourier transform is one of the most popular feature representation transforms for audio signal processing.~\cite{yoo2001tutorial}. In stationary signals where all frequency components exist throughout the entire duration, Fourier transform could extract all the required information from waves. However, for non-stationary signals, some frequency components do not appear during the entire span of the wave~\cite{polikar1996wavelet}. In order to deal with this problem, short-time Fourier transform (STFT) was developed to extract feature components assuming that the signal is relatively stable within a single short time window. Still, there is a trade-off between the resolution and temporal dynamics dependant on the window size.
\label{sec:pagestyle}
\subsection{Wavelet transformation}
The wavelet transform is developed as an alternative approach to the STFT to overcome the resolution problem~\cite{polikar1996wavelet}. A signal can be divided into wavelets by using the wavelet transform.
\begin{equation}
S_m= \frac{1}{{\vert a \vert}^{1/2}} \int_{-\infty}^{\infty}  x(t) \psi_m(\frac{t-b}{a}) dt
\end{equation}
where, $a \in R^{*+}$ is a scale and $b \in R$ is a translation value. $\psi_m$ is a wavelet basis. There could be a set of $\psi_m$. However, for ease of computation, people use orthonormal basis functions with the nice orthogonal properties. These basis functions include Morlet wavelet, Mexico Hat wavelet etc.
 
Also, to recover the original signal $x(t)$, an acoustic waveform can be represented by the inverse wavelet transform:
\begin{equation}
x(t) = <S_m, \psi_m(t)> \tilde{ \psi}_m(t)
\end{equation}
\begin{equation}
x(t) = C_{\psi}^{-1}  \int_{-\infty}^{\infty} \int_{-\infty}^{\infty}  S_m  \frac{1}{{\vert a \vert}^{1/2}} \tilde{\psi_m}(\frac{t-b}{a}) db \frac{db}{a^2} dt
\end{equation}
where $\tilde{\psi_m}$ is dual function of $\psi_m$, $C_{\psi}$ is a constant related with $\psi$.
These two equations could be further written as:
\begin{equation}
S_m = \sum x(t) \phi_m(t - \tau_m) 
\end{equation}
\begin{equation}
x(t) = \sum\sum S_m  \tilde{\phi_m}(t - \tau_m) 
\end{equation}
where, $\phi_m$ is a representation of the $\frac{1}{{\vert a \vert}^{1/2}} \psi_m(\frac{t-b}{a})$ and $\tilde{\phi_m}$ is a representation of the $\frac{1}{{\vert a \vert}^{1/2}} \tilde{\psi_m}(\frac{t-b}{a})$. $\tau_m$ is temporal shift of $\psi_m$. We can see that $S_m$ is derived from the convolution of original wave and the wavelet.

Note that in wavelet transform, the kernel $\phi_m$ is a single wavelet basis with different scale and translate. For a signal, various wavelet basises could be applied. and the choice of wavelet basis always leads to different performances. Meanwhile, instead of using a single basis, multiple wavelet basises could be applied.

Therefore, we hope to better understand the signal from the multiple waveform kernels learned through data-driven approaches. Convolutional neural network is one of such alternatives.

\subsection{Convolutional Layer of CNN}
Based on the understanding that the wavelet transform is the convolution of the raw waveform and a wavelet basis~\cite{holschneider1990real}, our hypothesis is that a convolutional layer have the same effect as wavelet transform.


Similar to the discrete convolution equation:
\begin{equation}
f(t)*g(t) = \sum_{-\infty}^{\infty} f(\tau) g(t - \tau)
\end{equation}
where $f(t)$ and $g(t)$ are two functions.
The output of the $m^{th}$ convolution unit of $i^{th}$ element is written as:
\begin{equation}
S_{i,m} = \sum_{j=i}^{i+f_1-1} w_{m,j-i}x_j + b_m
\end{equation}
where $w_{m,j-i}$ is $j-i$ element of the $m_{th}$ kernel. $f_1$ is the length of the kernel. $b_m$ is the $m_{th}$ kernel's bias. 
The output could then be further generalized as:
\begin{equation}
S_{m} = \sum w_{m}x + b_m
\end{equation}

It is shown that sound features are extracted by these convolutional layers where their kernals perform as filters. Both convolutional layers and wavelet transform have the capability of capturing signal's pattern with fine resolution while saving temporal dependency. 

Here, It is important to distinguish the difference between the wavelet transform and a convolutional layer. In wavelet transform, $\psi$ is an orthogonal wavelet basis. These basises are relatively hard to hand-craft due to the complex requirement. In contrast, for the convolutional layer, the learned kernel is dependant on the specific dataset with no need of hand-crafting. In addition, the representation of wavelet transform is strictly linear, while the inference of optimal kernel of the convolutional layer is highly non-linear and computational complex. Due to the high non-linearity, the original signal cannot be perfectly recovered from the convolutional layer. However, the pattern of the original waveform is captured through the abstraction of CNN.

\section{Experiment}
\label{sec:exp}
In this study, the training of the CNN model is performed on the natural sounds dataset from the urbansound8k dataset~\cite{salamon2014dataset}. This dataset contains 8732 labeled sound excerpts ($\leq 4s$) of urban sounds from 10 classes: air conditioner, car horn, children playing, dog bark, drilling, engine idling, gun shot, jackhammer, siren, and street music. These data are divided into 10 folds. Due to the dimension of the raw signal and GPU memory limits(4s with 44.1kHz, where the input vector size is 176.4k ), we down sample the signal with sampling rate 8kHz (the vector size is 32k).

The architecture of the network is shown in Fig.1. We use the default Adam optimizer~\cite{kingma2014adam} with initial learning rate of 0.001. The learning rate decays every three epochs with decay rate as $0.1$. We train the network with 30 epochs. If the training loss stop decreasing for 3 continuous epochs, the training will be terminated. The learning rate update function is:
\begin{equation}
lr = lr/(1+decayrate)
\end{equation}
where, $lr$ is the learning rate.
\begin{figure}[htb]
\begin{minipage}[b]{1.0\linewidth}
  \centering
  \label{fig1}
  \centerline{\includegraphics[width=4.1cm]{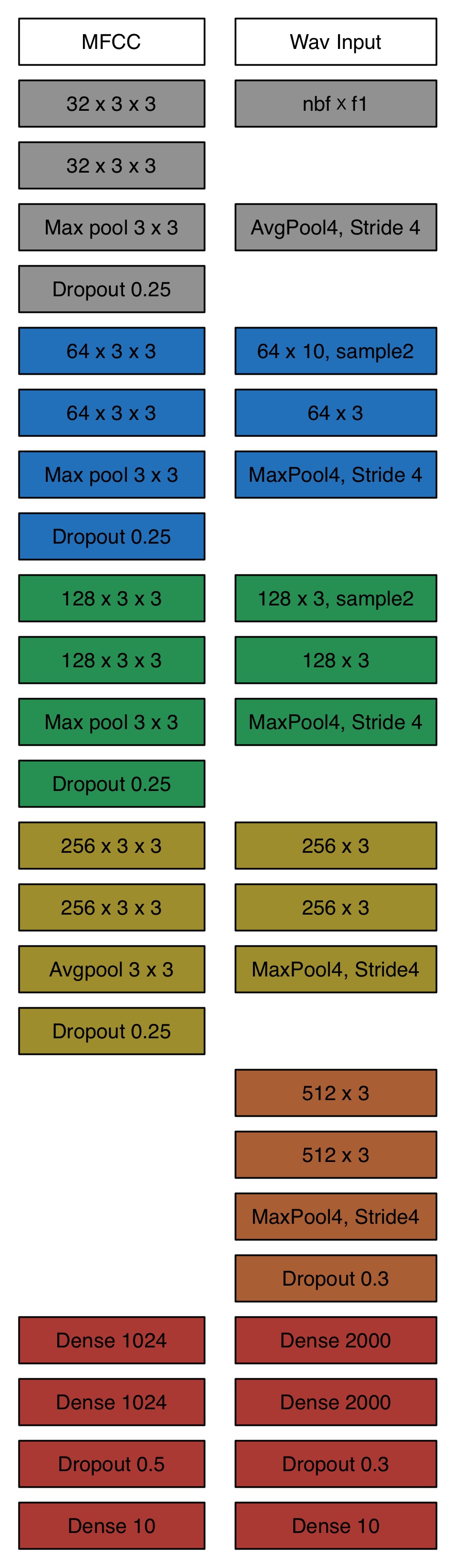}}
  \caption{We design two architectures. The left model is a deep VGG CNN that takes in MFCC feature(arch 1), the right one is a CNN structure that takes in raw waveform(arch 2)}\medskip
\end{minipage}
\end{figure}

In Fig.1, $f_1$ is the length of kernel, this has the similar interpretation as the window size of fft or wavelet transform. For instance, if $f_1 = 72$ and sampling rate equals to $8000Hz$, it means the window size is taken as $9ms$. $nb_f$ is the number of filters, which is similar to the number of fft factors or number of wavelets.

The baseline is around 70\%\cite{salamon2014dataset} by using svm with rbf kernel and 73.7\% \cite{piczak2015environmental}. 
\section{Result and Analysis}
\label{sec:resana}

\begin{table}[t]
\label{result}
\begin{center}
\begin{tabular}{llllll}
\multicolumn{1}{c}{\bf Pipeline} &
\multicolumn{1}{c}{\bf $f_1$} &
\multicolumn{1}{c}{\bf $nb_f$} &
\multicolumn{1}{c}{\bf n\_mfcc} &
\multicolumn{1}{c}{\bf Freq} &
\multicolumn{1}{c}{\bf Acc}
\\ \hline \\
Arch1 &72 &32 & &8    &65.97\\
Arch1 &160 &32 & &8    &62.12\\
Arch1 &72 &64 & &8    &63.53\\
Arch2 & &  &8  &40  &69.03\\
Arch2 & &  &22.5  &128  &78.34\\
\end{tabular}
\end{center}
\caption{Accuracy are shown in table1. The result is the average over ten folds. We clip the 4s audio into 4 1s clips, ensemble with majority voting for the last experiment.}
\end{table}

\subsection{Experiment Result}
\label{ssec:result}
The result is shown in Table 1. We first test with different window sizes. It is shown that compared with using a larger window of size 160 ($20ms$), the accuracy of using a 72 ($9ms$) window is 3\% better. We also change the number of filters. Increasing the number of filters to 64 does not help to improve the result. It is shown that during the training process, the loss keeps decreasing during the 30 epochs. Increasing number filters requires more training epochs to converge. By using MFCC as input feature, the accuracy reaches 69.03\%. We notice that the sampling rate of the sound affects the detection accuracy significantly, increasing the sampling rate to 22.5Hz greatly improves the accuracy. Therefore, relatively high sampling rate is essential for natural sound recognition tasks. However, for the raw input, it incurs a higher computational cost.

\subsection{Analysis}
\label{sssec:analysis}
The purpose of this experiment is to understand what does the first convolutional layer learned from the waveform, find the relation between wavelet transform and the convolution layer, and extract kernel as basis to better represent waveforms.

\subsubsection{kernel analysis of first convolutional layer}
Based on previous result, we further investigate the kernel learned by the first convolutional layer. Here, we take the folder 1's trained weight as an example. In~\cite{golik2015convolutional}, Fourier transform are performed on the weight matrix to estimate the bandwidth $f_b^{i}$ and the center frequency $f_c^{i}$. We first plot several power spectrum of weight in Fig.2. In the figure, each filter is activated when certain signal with corresponding frequency appears.
\begin{figure}[htb]
\begin{minipage}[b]{1.0\linewidth}
  \centering
  \label{fig2}
  \begin{tabular}{llll}
  {\includegraphics[width=2cm]{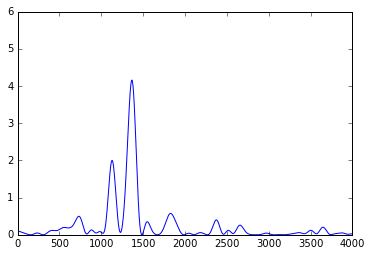}}
  {\includegraphics[width=2cm]{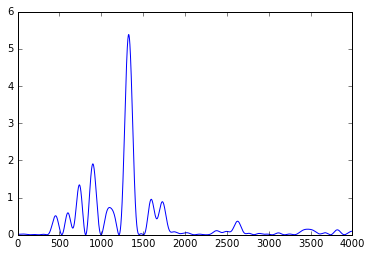}}
  {\includegraphics[width=2cm]{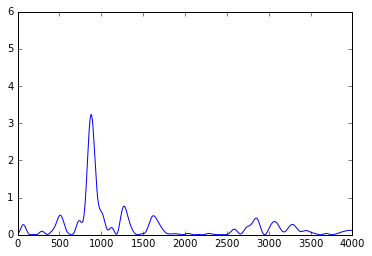}}
  {\includegraphics[width=2cm]{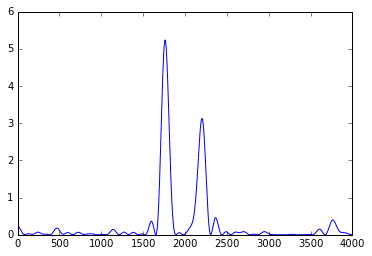}}
  \end{tabular}
  
  \caption{Several filters from the Conv1D layer. They are activated at different frequency range and perform as filters}\medskip
\end{minipage}
\end{figure}

Fig.3 (a) shows all filter's power spectrum, sorted by their center frequencies. The figure also shows that most filter's center frequency is below 2.5kHz and top few filters' center frequency are relatively unclear. Interestingly, from the Fig.3 (b), we can see that distribution the center frequencies are close to an exponential function. In wavelet transform, the scale parameter of a basis is also calculated by an exponential function. 

\begin{figure}[htb]
\begin{minipage}[b]{1.0\linewidth}
  \centering
  \label{fig3}
  \begin{tabular}{ll}
  {\includegraphics[width=3.5cm]{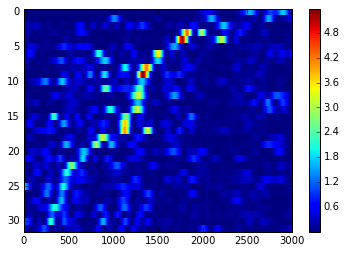}}
  {\includegraphics[width=3.5cm]{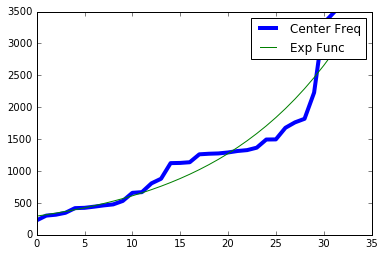}}
  \end{tabular}
  
  \caption{(a) spectrum of the filters, sorted by center frequencies. (b) plot of center frequencies.}\medskip
\end{minipage}
\end{figure}

We further investigate the convolutional layer's output by using a dog bark clip (101415-3-0-2.wav). In this clip, the dog barked at the very beginning and kept silent for a while as shown in Fig.4.

\begin{figure}[htb]
\begin{minipage}[b]{1.0\linewidth}
  \centering
  \label{fig4}
  \centerline{\includegraphics[width=8.5cm]{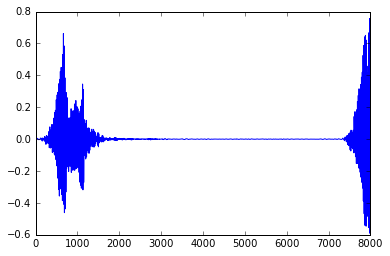}}
  \caption{Wav of dog bark}\medskip
\end{minipage}
\end{figure}

We feed the clip into the first convolutional layer and continuous wavelet transform(CWT). Both methods are activated when the dog start to bark as shown in Fig.5. In CWT, only certain area of first few filters fire, while most area of filters are activated in the convolutional layer.

\begin{figure}[htb]
\begin{minipage}[b]{1.0\linewidth}
  \centering
  \label{fig5}
  \begin{tabular}{ll}
  {\includegraphics[width=3.5cm]{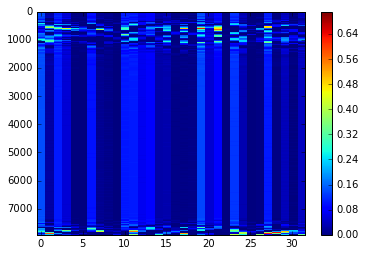}}
  {\includegraphics[width=3.5cm]{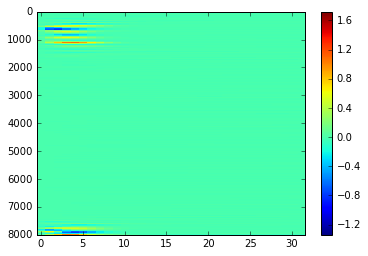}}
  \end{tabular}
  \caption{(a) output of convolutional layer (b) output of CWT }\medskip
\end{minipage}
\end{figure}

\subsubsection{Signal Recovery}
Both FFT and WT could be reversed and reconstruct original wave. Similarly, we try to reconstruct the origin signal by using the first convolutional layer. The reconstruction process is similar to reverse wavelet.
\begin{equation}
x(t) = \frac{1}{C_m}\sum (S_m - b_m) (w_m)^{-1}
\end{equation}
Here, we interpret the $w_m^{-1}$ as basis to encode the original waves generated through a data-drieven method. Fig.6 shows several smoothed basises from inverse weight of convolutional layer. These basises also show the asymmetric sinusoid~\cite{smith2006efficient}. The wave gradually decades along the time. 
\begin{figure}[htb]
\begin{minipage}[b]{1.0\linewidth}
  \centering
  \label{fig6}
  \begin{tabular}{lll}
  {\includegraphics[width=2.5cm]{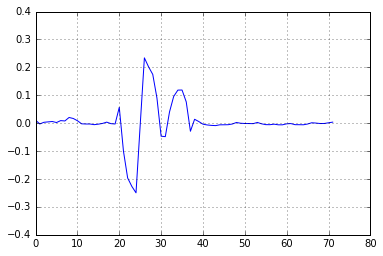}}
  {\includegraphics[width=2.5cm]{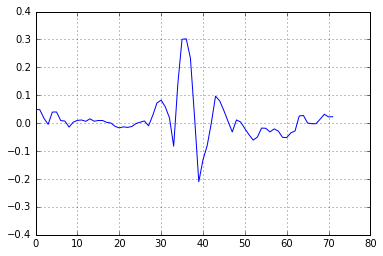}}
  {\includegraphics[width=2.5cm]{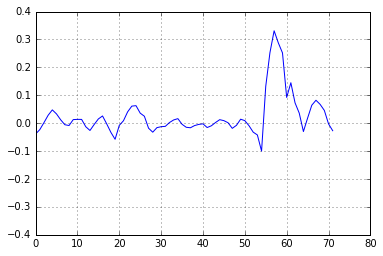}}
  \end{tabular}
  \caption{Coding basis}\medskip
\end{minipage}
\end{figure}

We take out the first piece of output on the first convolutional layer from dog bark clip, and recover the signal using eq.(10) by setting $C_m = 5.5$. We smooth and realign the recovered signal in Fig.7. The figure shows that the original signal is recovered with loss by using the output of the first convolutional layer. However, it still captures the general pattern of the original wave.
\begin{figure}[htb]
\begin{minipage}[b]{1.0\linewidth}
  \centering
  \label{fig7}
  \centerline{\includegraphics[width=8.5cm]{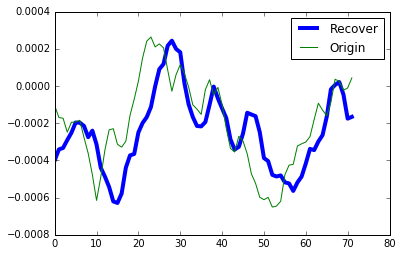}}
  \caption{Recovered Signal vs Original Signal}\medskip
\end{minipage}
\end{figure}

\section{Conclusion}
In this work, we extract features from a data-driven perspective by directly input raw waveforms into an end-to-end deep CNN architecture for sound recognition tasks. After training, we extract the weight from the first convolutional layer and apply Fourier transform on the weight, the result shows that the weight performs as bandpass filters. We also compare the CNN and wavelet transform. Finally, we reconstruct the signal by reversing the first convolutional layer. The recovered signal shows that the first convolutional layer could extract the pattern of original signal with minimum loss.

\bibliographystyle{IEEEbib}

\end{document}